# Atomic-Scale Visualization and Manipulation of Domain boundaries in 2D Ferroelectric In$_2$Se$_3$

*Fan Zhang[1‡], Zhe Wang[2‡], Lixuan Liu[3,4], Anmin Nie[4], Yongji Gong[3], Wenguang Zhu[2*], Chenggang Tao[1*]*

[1]Department of Physics, Virginia Tech, Blacksburg, Virginia 24061, United States.

[2]International Center for Quantum Design of Functional Materials (ICQD), Hefei National Laboratory for Physical Sciences at the Microscale, and Key Laboratory of Strongly-Coupled Quantum Matter Physics, Chinese Academy of Sciences, School of Physical Sciences, University of Science and Technology of China, Hefei, Anhui 230026, China.

[3]School of Materials Science and Engineering, Beihang University, Beijing 100191, China.

[4]Center for High Pressure Science, State Key Laboratory of Metastable Materials Science and Technology, Yanshan University, Qinghuangdao 066004, China.

**ABSTRACT**

**Domain boundaries in ferroelectric materials exhibit rich and diverse physical properties distinct from their parent materials and have been proposed for novel applications in nanoelectronics and quantum information technology. Due to their complexity and diversity, the internal atomic and electronic structure of domain boundaries that governs the electronic properties as well as the kinetics of domain switching remains far from being elucidated. By using scanning tunneling microscopy and spectroscopy (STM/S) combined with density functional theory (DFT) calculations, we directly visualize the atomic structure of domain boundaries in two-dimensional (2D) ferroelectric $\beta'$ In$_2$Se$_3$ down to the monolayer limit and reveal a double-barrier energy potential of the 60° tail to tail domain boundaries for the first time. We further controllably manipulate the domain boundaries with atomic precision by STM and show that the movements of domain boundaries can be driven by the electric field from an STM tip and proceed by the collective shifting of atoms at the domain boundaries. The results will deepen our understanding of domain boundaries in 2D ferroelectric materials and stimulate innovative applications of these materials.**

Domain boundaries are interfaces inside a material between neighboring domains of different crystallographic orientations. Due to symmetry breaking or topological protection, domain boundaries can be significantly different from their parent materials in various properties. A specific type of domain boundaries are polar domain boundaries that separate different regions of uniformly polarized domains, such as domain boundaries in ferroelectric materials. These polar domain boundaries have broad potential applications in nanoelectronics and quantum information technology, and thus have been intensively investigated.[1-12] However, microscopic understanding of domain boundaries in ferroelectric materials is still lacking because it is challenging to access the nanoscale domain boundaries that lie between adjacent domains with complex local interactions in the polarization and other structural parameters.[1, 13] Further, it is more challenging to control and manipulate the domain boundaries with atomic precision. Most previous atomic-scale studies of ferroelectric domain boundaries were conducted by transmission electron microscopy (TEM) that was able to resolve the structures of the domain boundaries with atomic resolution,[14, 15] but the local electronic structure of the internal domain boundaries remains elusive. As a powerful characterization tool with atomic resolution, scanning tunneling microscopy (STM) is capable of revealing both atomic structures and local electronic properties. Due to the difficulty to prepare cross-section STM samples of bulk ferroelectric materials and remove impurities that are energetically favorable to accumulate at the domain boundaries, STM studies on domain boundaries in bulk ferroelectric materials such as $BiFeO_3$ have been proven very challenging.[16, 17] Given the lower dimensionality, emerging two-dimensional (2D) ferroelectric materials[18-23] provide a new and exciting platform for exploring polar domain boundaries where the domain boundaries are relatively easy to access.

As a newly discovered 2D ferroelectric material, atomically thin $In_2Se_3$ has attracted numerous research efforts.[24-28] Particularly, 2D $In_2Se_3$ synthesized through the chemical vapor deposition (CVD) method is ideal for investigating the intriguing properties of polar domain boundaries, as a large amount of domain boundaries naturally form during the synthesis processes. Using scanning tunneling microscopy and spectroscopy (STM/S) combined with density functional theory (DFT) calculations, we directly visualize various types of domain boundaries in $β'$ $In_2Se_3$ down to the monolayer limit and resolve the atomic structures of the domain boundaries. Our STS measurements reveal a double-barrier structure with a width of about 3 nm in the 60° tail-to-tail domain boundaries in monolayer $\beta'$ In$_2$Se$_3$. The local density of states (LDOS) from the DFT calculations agree well with our experimental results. We further demonstrate the capability to manipulate the domain boundaries by STM with atomic precision. The results on elucidating the internal structure of domain boundaries in 2D ferroelectric materials and manipulating the domain boundaries at the atomic scale offer new opportunities for both fundamental studies of ferroelectric domain boundaries and their applications in data storage and electronic devices that take advantage of nanoscale ferroelectricity and localized functional properties.

**RESULTS AND DISCUSSION**

**Various types of domain boundaries in 2D $\beta'$ In$_2$Se$_3$**

Atomically thin In$_2$Se$_3$ films were grown on highly oriented pyrolytic graphite (HOPG) through the chemical vapor deposition (CVD) method.[25] At 77 K, In$_2$Se$_3$ is stabilized in the $\beta'$ phase as we previously reported.[25] The atomic model of the $\beta'$ In$_2$Se$_3$ structure is illustrated in **Figure 1**a and an atomically-resolved STM image of monolayer $\beta'$ In$_2$Se$_3$ is shown in Figure 1b, in which the unit cell is outlined by the black rectangle. As highlighted by the dashed rectangle in the side view model in Figure 1a, the surface Se atoms are at different heights, with the highest ones at the corner. The model is in excellent consistent with the experimental observation (Figure 1b), in which the surface Se atoms in a unit cell are overlapped on the STM image, with shades of color indicating the relative heights of Se atoms, dark blue represents higher while light blue represents lower. Two central Se atoms inside the unit cell arrange in a way that one is higher than the other. Theoretical calculations show that there is an in-plane polarization along the a axis while perpendicular to the b axis, pointing from the higher central atom to the lower one as marked by the green arrows. Because the $\beta'$ phase originates from the $\beta$ phase with reduced symmetry,[25] we expect various domain boundaries form between domains with different relative rotational angles in the $\beta'$ phase.

A typical large scale STM topography image of $\beta'$ In$_2$Se$_3$ surface is shown in Figure 1c, showing polycrystalline nature. Different domains can be distinguished by the stripes formed along the b axis of the $\beta'$ phase, which are marked by the blue dashed lines. A distinctive feature in this large scale STM images is the domain boundary, marked by the dashed green line. The domain boundary is nearly straight, along the bisector of the angle formed by the b axis of two neighboring domains. Here we define the domain boundaries by the angle between the b axes of neighboring domains, so the domain boundary in Figure 1c is a 60° domain boundary. From high resolution images of each domain as shown in the insets, we can further determine the polarization of each domain. The relationship of the polarization on each side is tail to tail in Figure 1c, so it is defined as a 60° tail to tail domain boundary. In Figure 1c the yellow region is bilayer In$_2$Se$_3$ and the black region corresponds to monolayer In$_2$Se$_3$. The left inset is the line profile along the marked blue line showing that the measured thickness of a single In$_2$Se$_3$ layer is 0.95 nm, same as previous reported results.[25, 29] The number of layers can be determined by measuring the height of each layer in an In$_2$Se$_3$ flake with respect to the HOPG substrate (see Figure S1). Figure 1d shows a monolayer $\beta'$ In$_2$Se$_3$, in which the 60° domain boundary on the left side in Figure 1d is a 60° head to tail domain boundary.

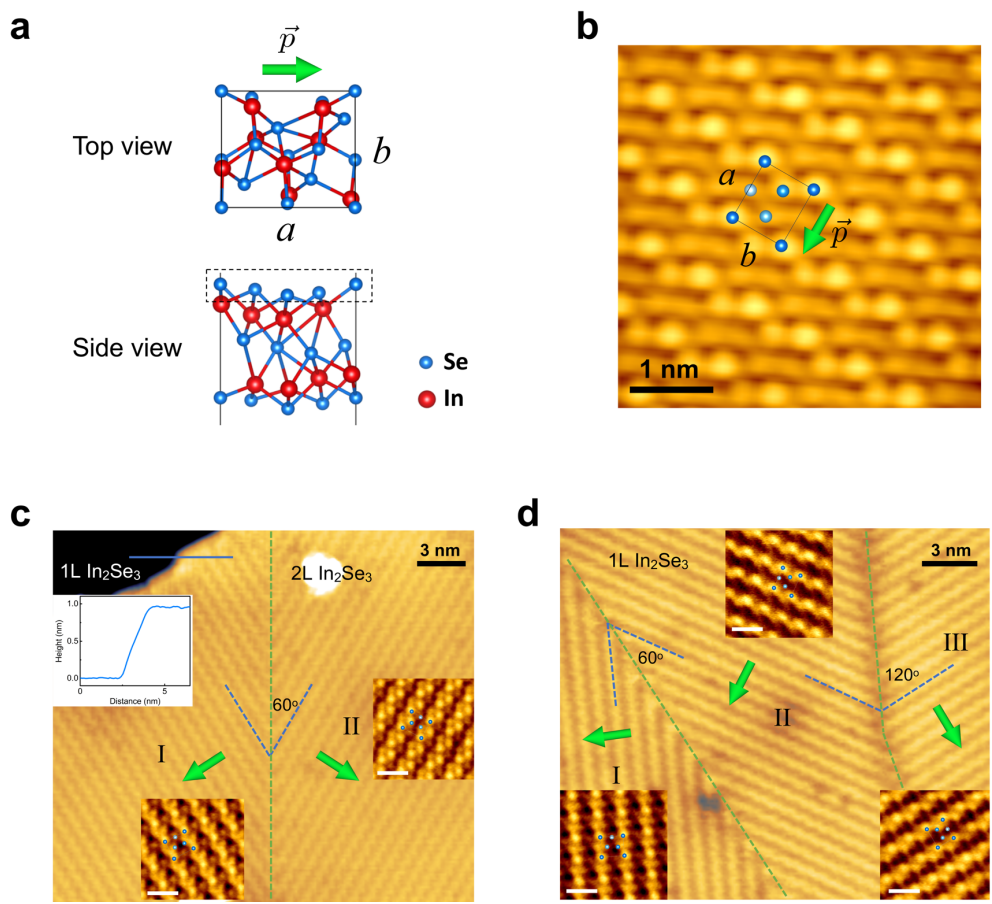

**Figure 1. Various types of domain boundaries in 2D $\beta'$ In$_2$Se$_3$.** **(a)** Atomic model of top and side view of $\beta'$ In$_2$Se$_3$ structure. **(b)** An atomically-resolved STM image of monolayer $\beta'$ In$_2$Se$_3$ ($V_S$ = 1.0 V, I = 0.4 nA). A unit cell is overlapped with the surface Se atoms from the atomic model, marked by the dashed rectangle in the side view in (a). The different brightness indicates the relative height of each Se atom. **(c)** Large scale STM image showing a 60° tail to tail domain boundary in bilayer In$_2$Se$_3$ ($V_S$ = 2.2 V, I = 0.16 nA). The left-up inset is the line profile along the blue line, showing the thickness of a single layer of $\beta'$ In$_2$Se$_3$. The high resolution STM images of domains I and II are shown as the insets at the middle-bottom and right respectively ($V_S$ = 1.25 V, I = 0.3 nA). The scale bars are 1 nm. **(d)** Large scale STM images showing 60° and 120° domain

boundaries in monolayer In$_2$Se$_3$ (V$_S$ = 2.0 V, I = 0.2 nA). The insets are high resolution images of each domain (V$_S$ = 1.3 V, I = 0.3 nA). The scale bars are 1 nm.

In addition to the 60° domain boundaries, 120° domain boundaries were also observed (Figure 1d), in which the boundary on the right side is a 120° domain boundary. In our experiments, both 60° and 120° domain boundaries were observed in monolayer and multilayer $\beta'$ In$_2$Se$_3$. Around 72% of all the observed domain boundaries are the 60° domain boundaries, and 27% are the 120° domain boundaries. Regardless of the polarization, the 60° domain boundaries are usually straight as shown in Figures 1c and 1d, while the 120° domain boundaries typically appear a little bit wandering, like the one in Figure 1d, indicating that the 60° domain boundaries are more stable at 77 K. As previously reported, 2D In$_2$Se$_3$ transforms from $\beta$ to $\beta'$ phase when temperature goes below around 180 K.[25] The $\beta'$ phase can form in six equivalent orientations inheriting from the three-fold symmetry of the $\beta$ phase (for details see Figure S2). Thus, 60°, 120° and 180° domain boundaries are expected to form. As schematically drawn in Figure S3, all the possible configurations for the 60° boundary and the 120° boundary are respectively listed. In our experiments, the majority domain boundaries observed are 60° and 120° types, while the 180° boundaries are rarely observed (see Figure S4). This may attribute to the high formation energy of the 180° boundaries.

**60° tail to tail domain boundaries in monolayer $\beta'$ In$_2$Se$_3$**

To explore the detailed structural and electronic properties, we now focus on the 60° domain boundaries in monolayer $\beta'$ In$_2$Se$_3$. **Figure 2**a is an atomically resolved STM image of a 60° tail to tail domain boundary in monolayer $\beta'$ In$_2$Se$_3$ obtained at a sample bias of 1.8 V. At this scale, we are able to clearly visualize the atomic structure of both the domain boundary and the vicinity area. The two neighboring domains on each side appear to connect seamlessly together, indicating the two domains originate from a same $\beta$ phase domain as described above. Along the domain boundary, a periodic structure is formed with a period of four atoms. The surface atomic structure beyond 2 to 3 unit cells away from the domain boundary on each side appears the same as bulk monolayer $\beta'$ In$_2$Se$_3$ and the polarization of each domain is determined from the relative heights of the central Se atoms in the unit cells in each domain away from the domain boundary. The STM topography images of domain boundary appear slightly different under various sample biases. Figures 2b and 2c are the empty and filled state images of the same area of Figure 2a, obtained at 0.7 V and –2.8 V sample bias respectively. At a sample bias of 1.8 V, the vicinity area of the domain boundary appears flat, while at a low positive sample bias (Figure 2b), typically less than 1.0 V, the topography of the domain boundary shows some characteristic features that two sides of the domain boundary are slightly dimmer, like two trenches along the domain boundary. The similar features also appear clearly in the filled state STM image (Figure 2c), in which the domain boundary appears higher and wider in comparison with Figure 2b (see also the line profiles in Figure S5).

To understand the atomic structure of domain boundaries, we turn to DFT calculations. A supercell with a 60° tail to tail domain boundary was constructed to simulate the experimentally observed structure (Figure S6). We then did the relaxation of all the atoms' positions of the initial structures to obtain the most stable structures. Figure 2d shows the fully relaxed atomic structure of a 60° tail to tail domain boundary. The rectangle on each side of the domain boundary in the top view (the top panel) marks the unit cell of $\beta'$ In$_2$Se$_3$ in each domain. For both the top and side views, the dashed line marks the center of the domain boundary. To compare with the experimental result,

Figure 2e superimposes the surface Se atoms of the optimized structure overlapped with the STM image of the area marked by the dashed rectangle in Figure 2a. It can be seen that the calculated result perfectly matches the measured atomic structure. Looking into the details of the optimized structure of the domain boundary, it reveals that the deviation of the atoms' positions from the unrelaxed positions of the initial structure, or called reference positions, mainly occurs within 1 to 2 unit cells around the center. To further clarify the displacement of atoms near the center of the domain boundary, we now focus on the Se atoms on the surface layer that are visualized in STM and the Se atoms in the central layer that determine the polarization of $\beta'$ $In_2Se_3$. As shown in Figure 2f, the top illustration reveals the displacement of the surface Se atoms and the bottom one for the Se atoms in the central layer, in which the blue dots represent the relaxed positions while the purple dots represent the reference positions. For the Se atoms of both the surface layer and the central layer, the displacements mainly occur within 1 to 2 unit cells around the center of the domain boundary (for details see Figure S7). It is worth noticing that the atomic displacement on two sides of the domain boundary center is slightly asymmetric. Considering the b vectors of the unit cells on two sides that are opposite to each other (see Figure S3), in principle there are two types of 60° tail to tail domain boundaries that are mirror-symmetric to each other with respect to the center of the domain boundary. In our experiments, the mirror-symmetric type of the domain boundary in Figure 2a was indeed observed, which is also in excellent consistence with the calculated structure (for details see Figure S8).

Because the displacements in the positions of the Se atoms in the central layer approximately represent the change of polarization near the domain boundary (for details see Note S1), by comparing the optimized positions of the central Se layer with the reference positions we are able to qualitatively evaluate the change of polarization near the center of the domain boundary, as indicated by the red arrows in Figure 2f. The polarization near the center of the domain boundary can be obtained by superimposing the change of the displacement on the initial position of $\beta'$ In$_2$Se$_3$ unit cell, as shown by the green arrows in Figure 2f. To understand the transition of polarization across the domain boundary from one side to the other side, a model of a 180° domain boundary is built (for details see Figure S9 and Note S1). Near the center of the domain boundary, the polarization in the x direction is significantly reduced, slightly increased in the y direction, and basically unchanged in the z direction. This implies that the 180° tail to tail domain boundary is a Neel type domain boundary.[13, 30, 31] We speculate that the 60° tail to tail domain boundary should also be a Neel type.

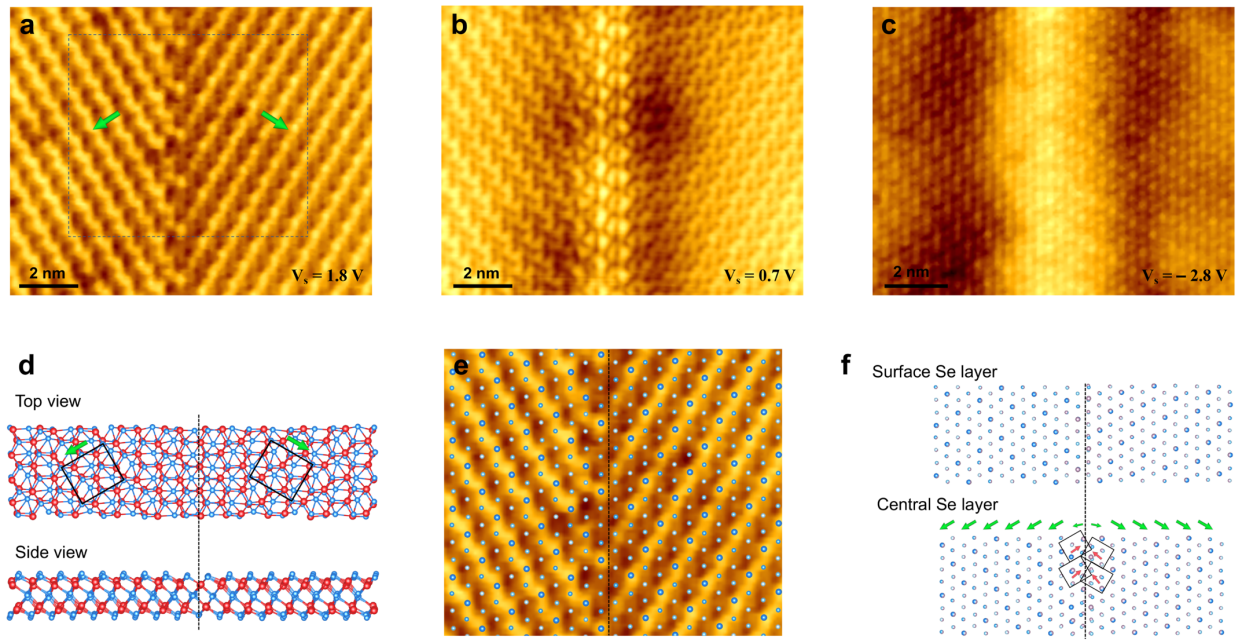

**Figure 2. STM images of a 60° tail to tail domain boundary in monolayer $\beta'$ In$_2$Se$_3$ and the calculated atomic model.** **(a)** An atomically resolved image of the 60° tail to tail boundary ($V_S$ = 1.8 V, I = 0.2 nA). The green arrows indicate the polarization direction in each domain. **(b)** Empty state image of the boundary obtained at a small positive sample bias ($V_S$ = 0.7 V, I = 0.16 nA). **(c)** Filled state image of the boundary ($V_S$ = −2.8 V, I = 0.16 nA). **(d)** The calculated atomic model of a 60° tail to tail boundary. **(e)** Zoom in STM image of the domain boundary overlapped with the top layer Se atoms from the calculated model. The area is marked by a dashed rectangle in (a). **(f)** Displacement of Se atoms in the surface and central layers. The blue atoms represent the relaxed Se atoms and the purple atoms represent the reference atoms. The green arrows indicate the direction and magnitude of the polarization as a function of the distance from the domain boundaries. The red arrows indicate the change of polarization of the unit cells in the relaxed atomic structure relative to the reference structure.

**Electronic structure of 60° tail-to-tail domain boundaries in monolayer $\beta'$ In$_2$Se$_3$**

To investigate the electronic structure of the domain boundaries at different energy levels, we conducted STS measurements across the 60° tail-to-tail domain boundaries (for details see Figure S8a). **Figure 3**a shows the dI/dV colormap as a function of bias and position, with the center of the domain boundary set as zero. The colormap can be roughly divided into two different regions based on the position, corresponding to the inside and outside of the domain boundary. The edges of the domain boundary regions can be defined as the transition positions as marked by the red and black arrows in Figure 3a, which are also approximately the locations of the trenches observed in the topography as shown in Figures 2a-c. From the definition of the edges, the width of the domain boundary is determined to be ~3 nm. Outside the domain boundary, both the valence and conduction band edges shift upward as the position moves closer to the center, a typical band bending scenario for the tail-to-tail domain boundaries in ferroelectric materials. Near the edge of the domain boundary, as marked by the black arrows in Figure 3a, the dI/dV density at the valence band edge becomes higher compared with the inside or outside regions. Inside the domain boundary, opposite to the outside region, the conduction band edge shifts downward toward the Fermi level when moving to the center of the domain boundary. Around the center of the domain boundary, faint in-gap states are also observed. Overall, the edges of the domain boundary exhibit electron confinement and behavior like a double-wall energy barrier separating the core of the domain boundary and outside regions. Outside the domain boundary, the bandgap of STS around 2.6 nm away from the center of the boundary is around 2.5 V, with the valence band edge at −1.9 V and the conduction band edge at 0.6 V, which is the same as the previous reported bandgap for bulk monolayer $\beta'$ $In_2Se_3$.[25] For STS near the center of the domain boundary, the bandgap is around 2.05 V with the valence band edge at −1.65 V and the conduction band edge at 0.4 V. The STS results can also explain the characteristic trench feature in the topological images and how the topological images evolve with the sample bias because in the constant current topographic mode the tunneling current is dependent on both the tip sample separation and the integral of the density of states from Fermi level to the sample bias. Furthermore, the slight asymmetry of the two sides of the domain boundary is also visible in the STS across the domain boundary.

Using DFT, we reproduced the dI/dV colormap of Figure 3a *via* calculating the LDOS across a 60° tail to tail boundary as shown in Figure 3b. Its overall features agree quite well with the experimental results. In the outside regions of the domain boundary, both the conduction band and valence band bend upward as approaching the domain boundary, owing to the tail-to-tail in-plane polarization. While in the inside region of the domain boundary, some in-gap states appear at the center of the domain boundary, and the band edges of both the conduction band and valence band move downward. Such band shifts can be attributed to the variation of the band alignment with respect to the vacuum level as the atomic structure transforms across the domain boundary. As revealed in Figures 2d and 2f, at the center region of the domain boundary, the electric polarization nearly vanishes, and the atomic structure approaches the $\beta$ phase. The calculated band alignments of the *β* and *β'* phase, as shown in Figure S10, confirms the experimentally observed downward shift of the band edges of the *β* phase with respect to those of the *β'* phase.

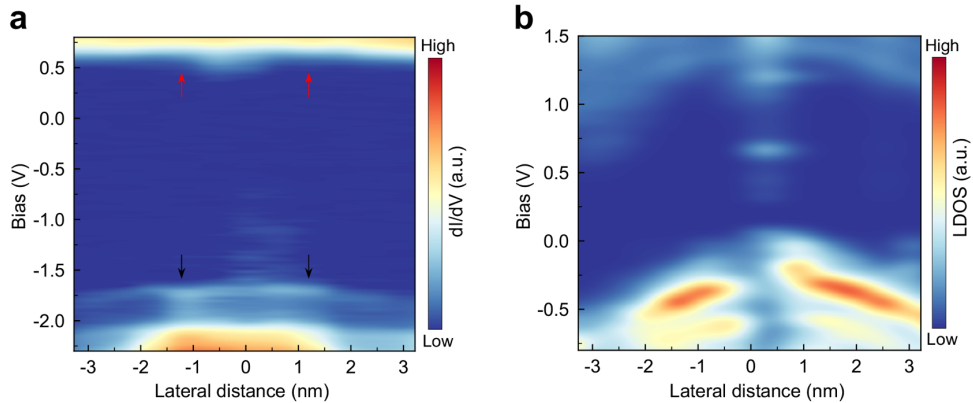

**Figure 3. Electronic structure of 60° tail-to-tail domain boundaries in monolayer *β'* In$_2$Se$_3$.** **(a)** dI/dV spectrum colormap across a 60° tail-to-tail domain boundary. All the spectra were acquired at the set point of $V_S$ = 1.0 V and I = 0.4 nA with the lock-in modulation of 20 mV. **(b)** Calculated LDOS across a 60° tail-to-tail domain boundary.

**Manipulation of a domain boundary in monolayer *β'* In$_2$Se$_3$ with atomic precision**

For their potential applications in data storage and quantum information science, it is essential to control and manipulate domain boundaries.[32-34] In our experiments, we further used STM to manipulate the domain boundaries. **Figure 4**a shows a 60° tail to tail domain boundary (see Figure S11 for details about determination of the domain boundary type) with the initial position marked by the dashed green line along the brightest atoms on the domain boundary. The defects can be used as reference points for determining the movement of the domain boundary, two of which highlighted by the dashed and solid circles in Figure 4a. Under a sample bias of 4.2 V and continuous scanning, the electric field from the STM tip is able to drive the domain boundary to move as shown in a series of STM images in Figures 4a-4c. Overall, the 60° domain boundary moves in a step-by-step manner from the initial position to new positions marked by a dashed red line in Figure 4c and a dashed blue line in Figure 4e, respectively. Although the domain boundary prefers a straight configuration, during the moving process, usually parts of the domain boundary move first and then the whole reaches the new position, as shown in Figures 4b and 4d. In most cases, the section moves to the left side as shown in Figures 4a to 4e. However, during the process, the opposite motion is also occasionally observed, implying a small difference between the energy barriers for transition of the domain boundary to the left or to the right.

Figures 4f-4h are zoom in images of Figures 4a, 4c and 4e, overlapped with the atomic models, in which the positions of the domain boundary are labeled as P1, P2 or P3. The domain boundary moves from one position to the nearest neighboring position while keeps the same configuration of the domain boundary. The step length between the nearest neighboring positions marked in Figures 4g and 4h is 2b, where b is the lattice constant along the b axis. It is worth noticing that the collective movement of the domain boundary can overcome some point defects, one of which marked by the solid circles in Figures 4a-4h. To understand the movement of the domain boundary, we calculated the kinetic pathways of switching the polarization as a simplified model to estimate the energy barrier for moving the domain boundary. As shown in Figure 4i, the initial and final states respectively correspond to the atomic structures of the left and right domains. As the domain boundary moves to the left, the initial state changes to the final state. The energy to switch the polarization orientation is 0.131 eV/unit cell. The overall biased motion, in this case from the right

to the left, should be related to the asymmetry in structural and electronic structures of the two sides of the domain boundary as described above.

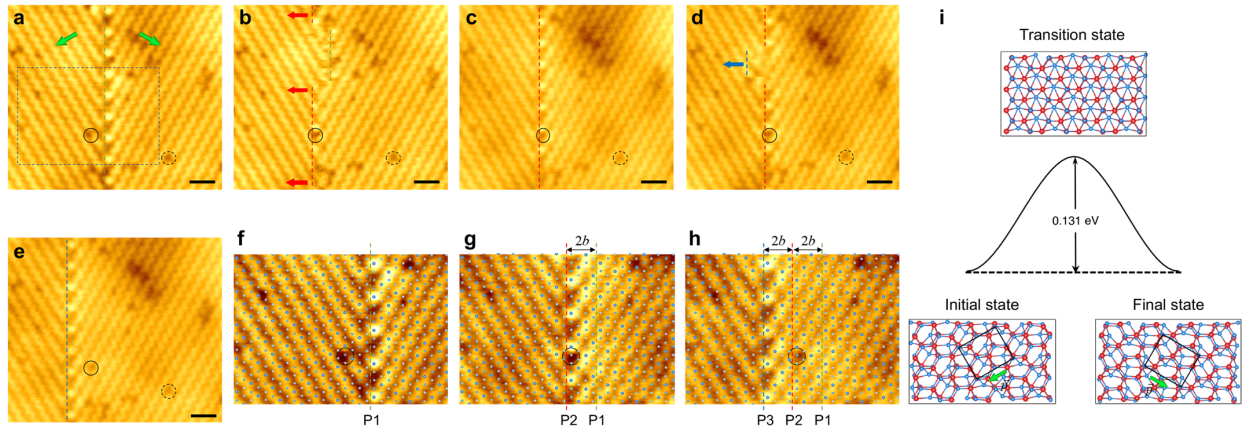

**Figure 4. Manipulation of a domain boundary in monolayer β′ In$_2$Se$_3$ with atomic precision.** **(a-e)** A series of STM images showing the motion of a 60° tail to tail domain boundary. The green arrows in (a) indicate the polarization direction of each domain. The initial position of the domain boundary is marked by the green dashed line in (a), the following two positions of the domain boundary are indicated by red and blue dashed lines in (c) and (e). (b) and (d) show parts of the domain boundary moved. Scanning parameters for (a-e): $V_S$ = 4.2 V, I = 0.2 nA, and Scanning speed = 600 s/image. All the scale bars are 2 nm. **(f-h)** Zoom in STM images overlapped with the atomic models of (a), (c) and (e), respectively. The area is marked by the dashed rectangle in (a). **(i)** calculated kinetic pathway and energy barrier of switching the polarization from the left domain to the right domain. The built supercell here is the same as the one for calculating the domain boundary (Figure S6).

**CONCLUSION**

In summary, we have directly visualized the internal atomic and electronic structure of domain boundaries in 2D ferroelectric $\beta'$ In$_2$Se$_3$ with thickness down to the monolayer limit by using STM/S in combination with DFT calculations. We reveal a double barrier structure with a width of about 3 nm of the 60° tail to tail domain boundaries in monolayer $\beta'$ In$_2$Se$_3$ from the STS crossing the domain boundaries. In addition, we demonstrate atomic scale manipulation of domain boundaries by STM and show that the movements of the 60° tail to tail domain boundaries in monolayer $\beta'$ In$_2$Se$_3$ can be driven by the electric field from an STM tip and proceed by the collective shifting of atoms at the domain boundaries. The results will deepen our understanding on domain boundaries in emerging 2D ferroelectric materials and pave a way for their broad applications in nanoelectronics and quantum information technologies.

## METHOD

### In$_2$Se$_3$ synthesis

The In$_2$Se$_3$ nanosheets were grown on highly oriented pyrolytic graphite (HOPG) substrates by a chemical vapor deposition method inside a homemade multiple temperature zone tube furnace. High pure selenium (Alfa Aesar, purity 99.999%) powder was evaporated at the upstream with a temperature stabilized at 270 °C, while In$_2$O$_3$ powder (Alfa Aesar, purity 99.99%) and HOPG substrates (SPI Supplies, USA) were placed at the upstream of quartz tube with temperature at 750 and 640 °C, respectively. The selenium and In$_2$O$_3$ vapors were carried to the HOPG substrate with a 20 sccm gas flow (Ar:H$_2$ ~4:1) controlled by a mass-flow controller.

### STM characterization

STM and STS characterizations were carried out in an ultra-high vacuum (UHV) STM system (a customized Omicron LT STM/Q-plus AFM system). After transferred into the preparation chamber of the STM system with a base pressure of < 10$^{-10}$ mbar, the samples were annealed at 380 °C for 2 hours in the chamber before transferred in vacuum to the STM chamber that is connected to the preparation chamber. An optical microscope (Infinity K2) mounted to one of the optical windows of the STM chamber was used to locate the STM tip to the desired areas on the samples. STM/STS measurements were then performed in the STM chamber at 77 K.

### DFT calculation

The density functional theory calculations were performed using the Vienna Ab *Initio* Simulation Package (VASP),[35] which is based on the plane wave basis sets with the projector-augmented wave (PAW) method.[36, 37] The exchange and correlation functional was treated using the Perdew-Burke-Ernzerhof (PBE)[38] parameterization of generalized gradient approximation (GGA). In order to model the 2D films, the supercells contain a vacuum region of more than 15 Å. A Γ-centered 6×7×1 and 2×3×1 k-point meshes were used for Brillion zone sampling for the unit cell and the built unit of $\beta'$ In$_2$Se$_3$, respectively. The gamma-only version of VASP was used for the supercell with domain boundary. The energy cut-off of the plane wave basis was set as 250 eV. Electronic minimization was performed with a tolerance of $10^{-4}$ eV, and ionic relaxation was performed with a force tolerance of 0.01 eV/Å on each ion. All these parameters were carefully tested to ensure the convergence and accuracy. The in-plane electric polarization was evaluated by using the Berry phase method.[39] The climbing image nudged elastic band method was used to determine the energy barriers of the kinetic processes.[40]

ASSOCIATED CONTENT

**Supporting Information**.

The Supporting Information is available free of charge.

Determination of the number of $\beta'$ In$_2$Se$_3$ layers, scanning induced phase transition at a mixed $\beta$ and $\beta'$ region in monolayer In$_2$Se$_3$, additional data of various types of domain boundaries, schematic illustration of the alignments of the conduction band minimum (CBM) and valence band maximum (VBM) of the $\beta'$ phase and the $\beta$ phase In$_2$Se$_3$ with respect to the vacuum level. (PDF)

**AUTHOR INFORMATION**

**Corresponding Author**

*E-mail: wgzhu@ustc.edu.cn;

*E-mail: cgtao@vt.edu.

**Author contributions**

C.T. conceived and designed the research. F.Z. and C.T. performed the experiments and analyzed the data. Z.W. and W.Z. carried out the theoretical calculations. F.Z. and Z.W. contributed equally to this work. L.L. and A.N. synthesized In$_2$Se$_3$. F.Z. and C.T. wrote the paper with support from Z. W. and W. Z.. All authors discussed the results and contributed to manuscript revisions. ‡These authors contribute equally to this work.

**NOTE**

The authors declare no competing interest.

**ACKNOWLEDGEMENTS**

F.Z. and C.T. acknowledge the financial support provided for this work by the U.S. Army Research Office under Grant W911NF-15-1-0414. L.L. and A.N. acknowledge support from the National Natural Science Foundation of China (Grant No. 51732010). Z.W. and W.Z. acknowledge National Key Research and Development Program of China (Grant No. 2017YFA0204904), the National Natural Science Foundation of China (Grant Nos. 11674299 and 11634011), the Fundamental Research Funds for the Central Universities (Grant Nos. WK2340000063, WK2340000082, and WK2060190084), the Strategic Priority Research Program of Chinese Academy of Sciences (Grant No. XDB30000000), and Anhui Initiative in Quantum Information Technologies.

Supporting Information for

# Atomic-Scale Visualization and Manipulation of Domain Boundaries in 2D Ferroelectric In$_2$Se$_3$

*Fan Zhang[1‡], Zhe Wang[2‡], Lixuan Liu[3,4], Anmin Nie[4], Yongji Gong[3], Wenguang Zhu[2*], Chenggang Tao[1*]*

[1]Department of Physics, Virginia Tech, Blacksburg, Virginia 24061, United States.

[2]International Center for Quantum Design of Functional Materials (ICQD), Hefei National Laboratory for Physical Sciences at the Microscale, and Key Laboratory of Strongly-Coupled Quantum Matter Physics, Chinese Academy of Sciences, School of Physical Sciences, University of Science and Technology of China, Hefei, Anhui 230026, China.

[3]School of Materials Science and Engineering, Beihang University, Beijing 100191, China.

[4]Center for High Pressure Science, State Key Laboratory of Metastable Materials Science and Technology, Yanshan University, Qinghuangdao 066004, China.

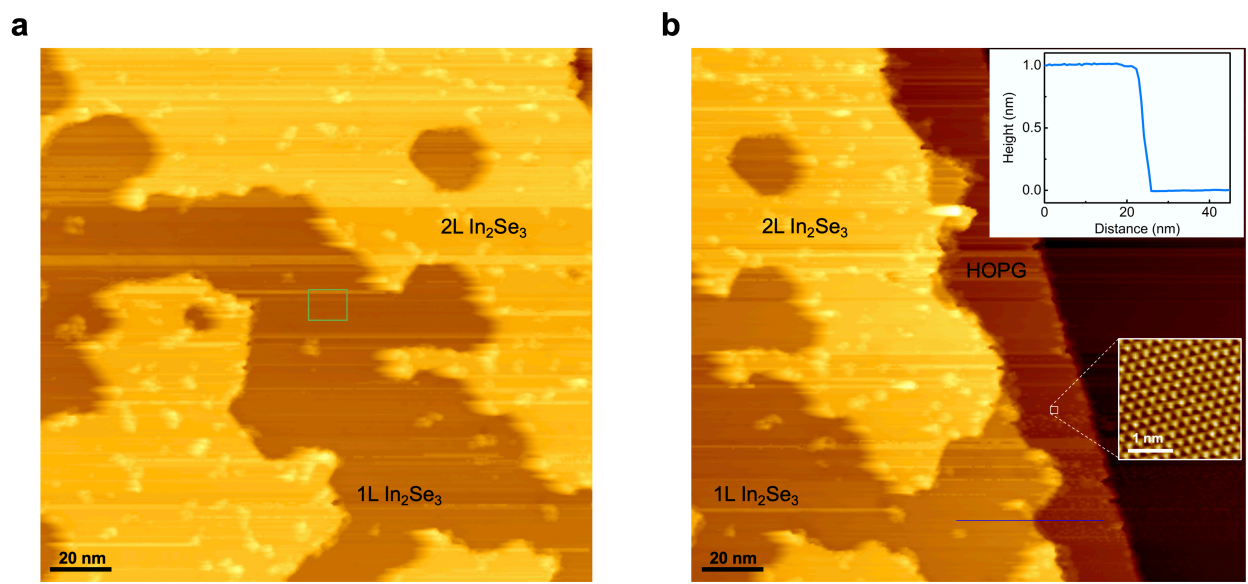

**Figure S1. Determination of the number of $\beta'$ In$_2$Se$_3$ layers. a**, A large scale STM image of monolayer and bilayer $\beta'$ In$_2$Se$_3$ ($V_S$ = 2.0 V, I = 0.2 nA). The green rectangle marks the area of Figure 2. **b**, A large scale STM image of monolayer and bilayer $\beta'$ In$_2$Se$_3$ around the edge of the In$_2$Se$_3$ flake ($V_S$ = 2.0 V, I = 0.2 nA). The top inset is the line profile obtained from the marked blue line, showing the thickness of a single $\beta'$ In$_2$Se$_3$ layer. The bottom inset is an atomically resolved STM image of the HOPG substrate obtained from area marked by the small white square ($V_S$ = − 0.1 V, I = 1.0 nA). The region in (b) is slightly to the right of the region in (a), with the left part of (b) same as the right part of (a).

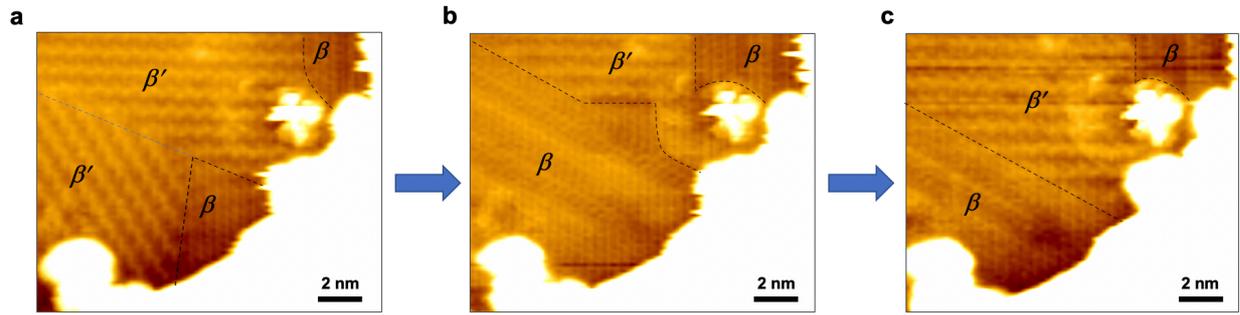

**Figure S2. Scanning induced phase transition at a mixed *β* and *β′* region in monolayer In₂Se₃. a**, **b** and **c** are sequential images under continuous scanning with the scanning parameters of $V_S$ = 2.0 V and I = 0.16 nA, showing the phase transition between the *β* and *β′* phases. The black dotted lines in the images indicate the phase boundary between the *β* and *β′* phases. The blue dotted line in (a) marks the domain boundary within the *β′* phase. The bright white area is bilayer *β′* In₂Se₃.

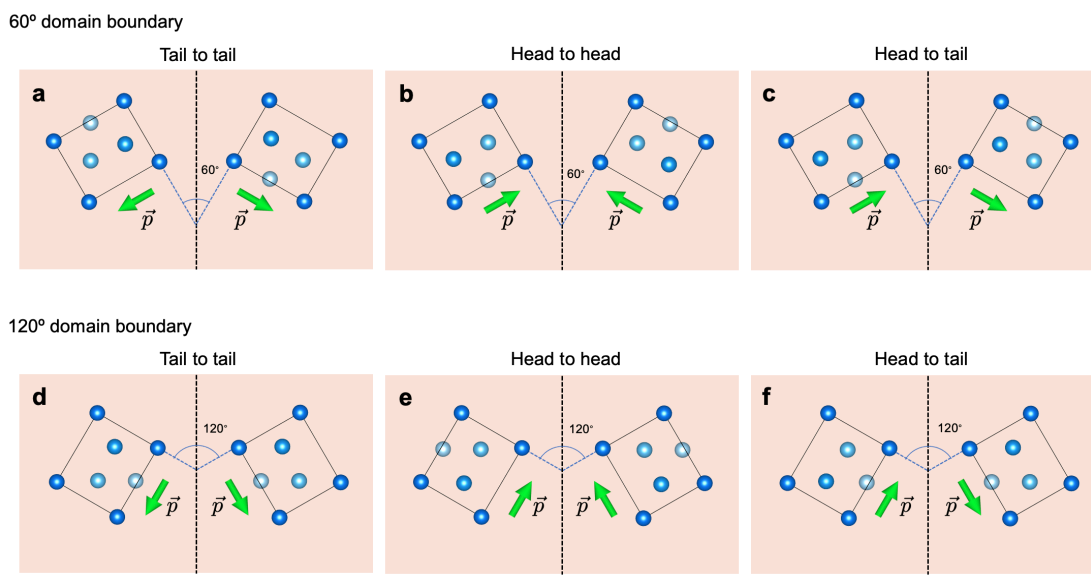

**Figure S3. Illustrations of various types of 60º and 120º domain boundaries.** The blue dashed lines are along the b directions of the unit cell in each domain, forming a 60º or 120º angle. The boundary serving as the angle bisector is marked with black dashed lines. Based on the polarization direction, there are three possible structures for 60º or 120º boundaries which are (**a-c**) 60º and (**d-f**) 120º tail to tail, head to head and head to tail domain boundaries, respectively.

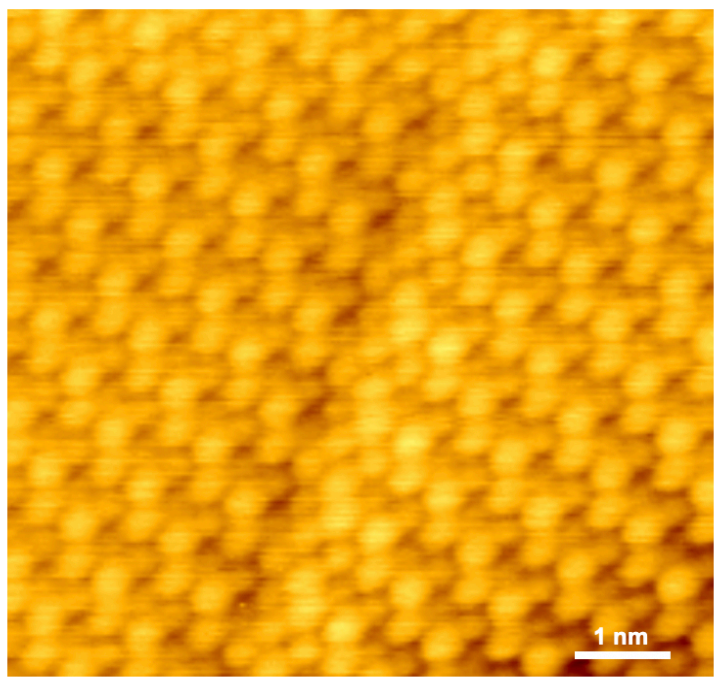

**Figure S4. A 180º domain boundary in multilayer $\beta'$In$_2$Se$_3$** ($V_S$ = 1.5 V, I = 0.4 nA).

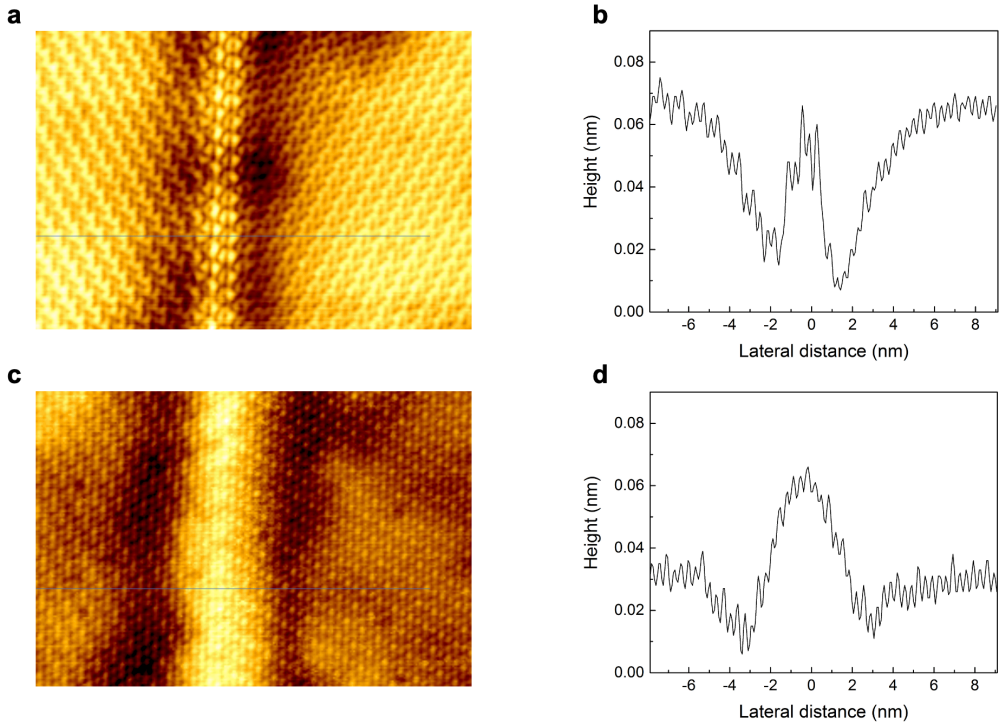

**Figure S5. Line profiles across a 60° tail to tail domain boundary in monolayer $\beta'$ In$_2$Se$_3$.** **a**, Figures 2b in the main text, an empty state image of the tail to tail domain boundary obtained at a small positive sample bias ($V_S$ = 0.7 V, I = 0.16 nA). **c,** Figures 2c in the main text, a filled state image of the boundary ($V_S$ = –2.8 V, I = 0.16 nA). **b,** and **d,** line profiles across the domain boundary along the marked lines marked in (a) and (c), respectively.

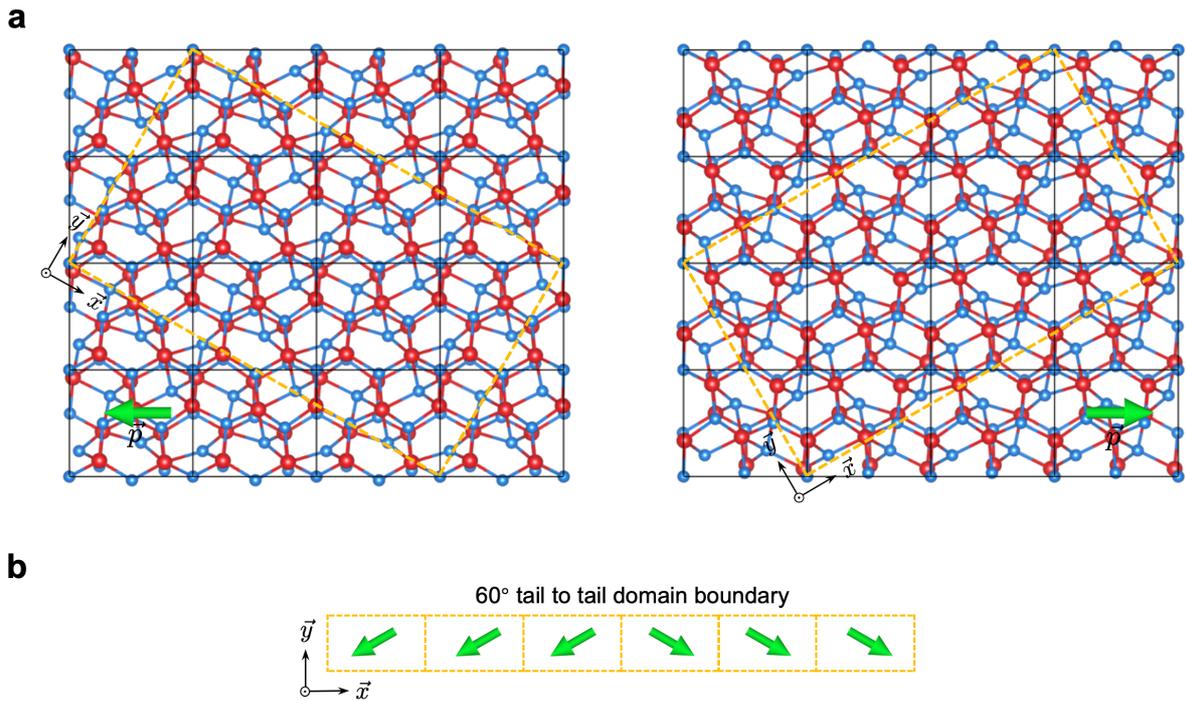

**Figure S6. Supercell structures of a 60° tail to tail domain boundary. a**, Unit cells of $\beta'$ In$_2$Se$_3$ (small black rectangles) with polarization along the horizontal direction. A built unit (orange rectangle) to simulate angled domain boundaries. **b**, Schematic of the geometry of the domain boundary structure with 60° as defined in the experiment. The left domain and the right domain are respectively from the left built unit and the right built unit in (a). Green arrows denote the polarization directions in all figures.

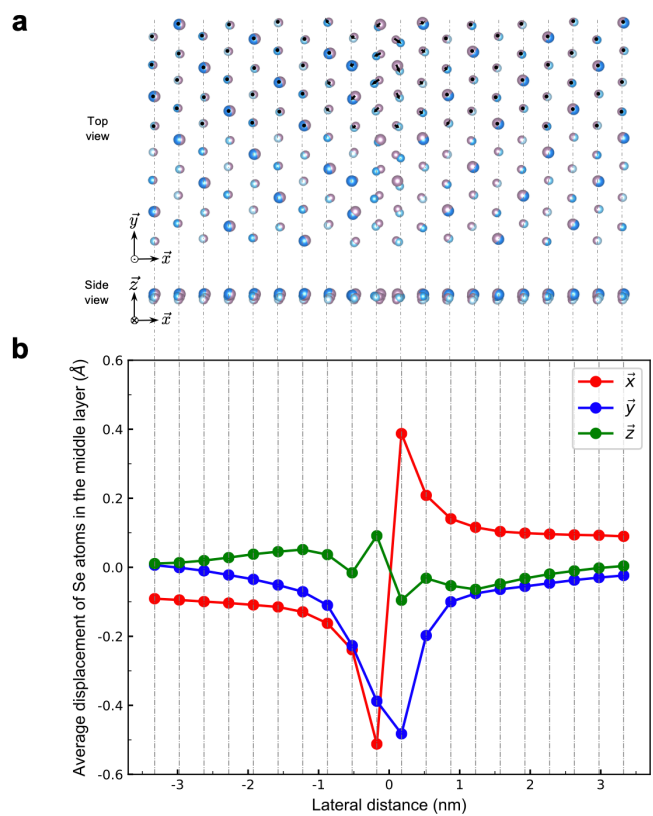

**Figure S7. Atomic displacements of the central Se layer in a 60° tail to tail domain boundary. a,** Top and side views of the atomic displacements of the central Se layer in a 60° tail to tail domain boundary. The structure is same as the one in Figure 2d in the main text. The optimized and reference positions are indicated by blue and light purple atoms, respectively. The displacements between the optimized and the reference positions are shown by the black arrows. **b,** The average displacements at x, y, z directions from the reference lattice structure as a function of lateral distance from the center of the domain boundary.

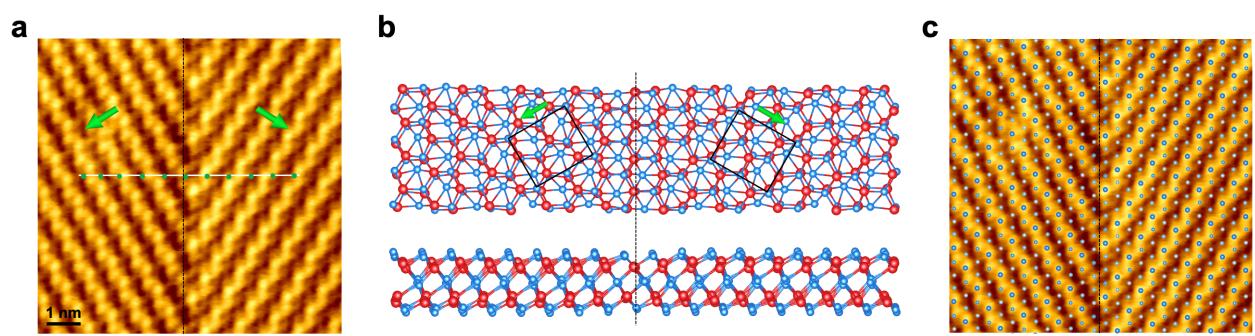

**Figure S8. The mirror-symmetric type of the 60° tail to tail domain boundary. a**, STM image of a 60° tail-to-tail domain boundary in monolayer In$_2$Se$_3$ ($V_S$ = 1.7 V, I = 0.4 nA). The green arrows indicate the polarization direction in each domain. The dI/dV spectra in Figure 3a were acquired at the green points along the dashed white line (from right to left). **b**, The calculated atomic models of the domain boundary, which is the mirror-symmetric type of the 60 tail to tail domain boundary shown in Figure 2 in the main text. **c**, The domain boundary shown in (a) overlapped with the surface Se atom in the calculated model. The black dotted lines in (a), (b) and (c) mark the center of the domain boundary.

**Note S1:** The supercell with a 180° tail to tail domain boundary consists of 16×1×1 $\beta'$ In$_2$Se$_3$ unit cell stacked in the x direction, where the polarization of the left half points to left and the right half to right. The top panel of Figure S9 only shows partial atomic structure of the supercell close to the tail to tail domain boundary. Here, we use the number R1 unit cell as an example to illustrate the relationship between atomic displacements and the change of polarization. Comparing the polarization of the number R1 unit cell with that away from the domain boundary (such as number R3, R4, R5) in the bottom panel of Figure S9, it is obvious that the polarization in the x direction is significantly reduced, slightly increased in the y direction, and basically unchanged in the z direction. Therefore, we focus on the in-plane atomic displacements. The larger top view of the number R1 unit cell is shown in the middle panel of Figure S9. The in-plane positions of In atoms are almost unchanged, while the displacements of Se atoms are relatively obvious. Comparing with the reference positions (purple atoms), the relaxed positions (blue atoms) of Se atoms moved rightward or lower rightward, resulting in a dipole moment to the upper left for the unit cell, as shown in the red arrow in the larger top view of the number R1 unit cell. There is a rightward polarization for the reference unit cell. The final polarization of the number R1 unit cell can be obtained by superimposing the upper leftward dipole moment on the rightward polarization (the initial polarization of $\beta'$ In$_2$Se$_3$ unit cell), as shown by the green arrows above the larger top view of the number R1 unit cell. The leftward (upward) dipole moment qualitatively analyzed by the displacements of the central Se atoms is consistent with the change of polarization, decreases in the x direction (increases in the y direction), quantitatively calculated by the Berry phase method. Therefore, the displacements in the positions of the Se atoms in the central layer can approximately represent the change of polarization near the boundary, which can be used to analyze the other angle domain boundaries easily.

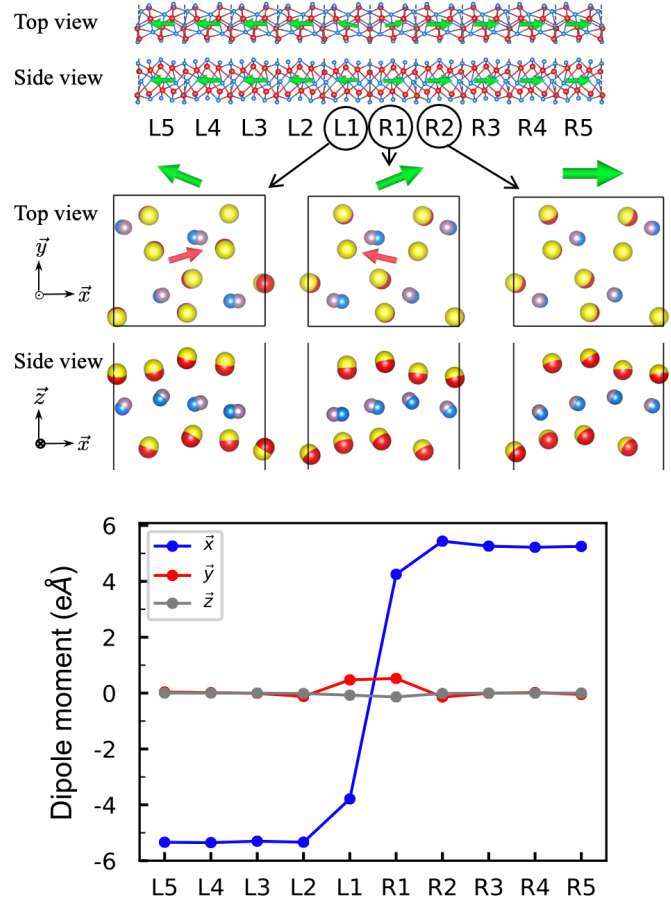

**Figure S9. Polarization inversion near a 180° tail to tail domain boundaries. Top**, The atomic structure of the domain boundary in top and side views is shown underneath the position axis, where green arrows denote the polarization directions per unit cell. **Middle**, The central In-Se-In layer atoms of unit cells near the boundary (L1, R1, and R2) are specifically zoomed in to show the detailed displacements of atoms. The light purple and blue atoms respectively show the positions of Se atoms before and after relaxation. The yellow and red atoms respectively show the positions of In atoms before and after relaxation. The red arrows indicate the change of polarization of the unit cells in the relaxed atomic structure relative to the reference structure. **Bottom**, The dipole moment of each unit cell in the x, y, and z direction are indicated by blue, red, and grey, respectively.

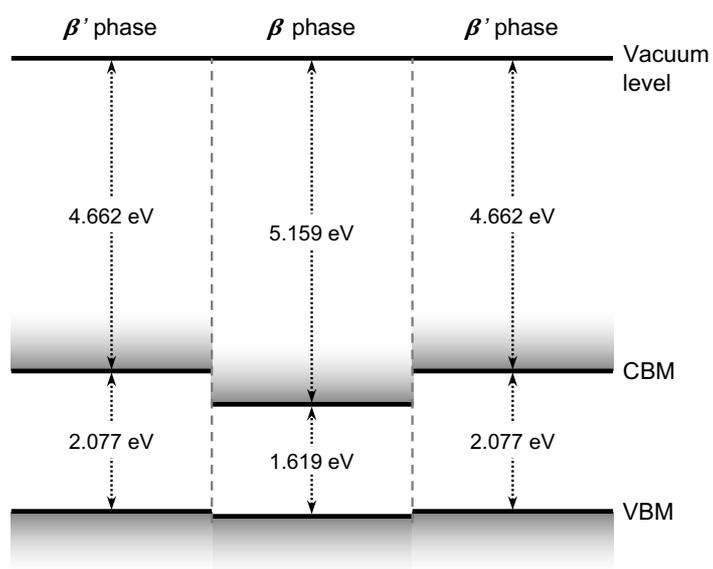

**Figure S10.** Schematic illustration of the alignments of the conduction band minimum (CBM) and valence band maximum (VBM) of the *β′* phase and the *β* phase In$_2$Se$_3$ with respect to the vacuum level.

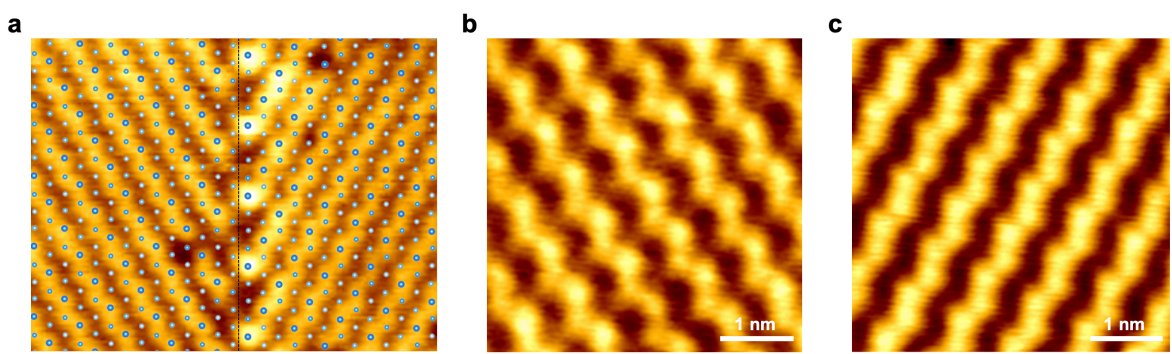

**Figure S11. High resolution STM images of the domain boundary and each domain in Figure 4 in the main text. a**, The domain boundary overlapped with top layer Se atoms from the calculated structure ($V_S$ = 4.2 V, I = 0.2 nA). **b**, **c**, High resolution STM images obtained respectively from the left (b) and right (c) domain of the domain boundary. Scanning parameters for both (b) and (c): $V_S$ = 1.8 V, I = 0.4 nA.

13